# Practical Crystallography with a Transmission Electron Microscope

Benjamin L Weare,[1,2] Kayleigh L Y Fung,[3*] Ian Cardillo-Zallo,[1,4] William J Cull,[2†‡] Michael W Fay,[1,5] Stephen P Argent,[2] Paul D Brown[5]

1 - Nanoscale and Microscale Research Centre, University of Nottingham, Nottingham NG7 2RD, United Kingdom.

2 - School of Chemistry, University of Nottingham, Nottingham NG7 2RD, United Kingdom.

3 - Department of Computer Science, Nottingham Trent University, Nottingham NG11 8NS, United Kingdom.

4 – School of Pharmacy, University of Nottingham, Nottingham NG7 2RD, United Kingdom.

5 - Department of Mechanical, Materials, & Manufacturing Engineering, Faculty of Engineering, University of Nottingham, Nottingham NG7 2RD, United Kingdom.

Corresponding author email: benjamin.weare1@nottingham.ac.uk

[*] Bruker UK, Electric Works, 3 Concourse Way, Sheffield, S1 2BJ, United Kingdom.

[†] SuperSTEM, SciTech Daresbury Science and Innovation Campus, Daresbury, WA4 4AD, United Kingdom.

[‡] School of Physics, Engineering, and Technology, University of York, York, YO10 5DD, United Kingdom.

**Abstract**

Three-dimensional electron diffraction (3DED) is a powerful technique providing for crystal structure solutions of sub-micron sized crystals too small for structure determination via X-ray techniques. The entry requirement, however, of a transmission electron microscope (TEM) adapted with bespoke software for coordinated sample stage rotation and continuous electron diffraction data acquisition has generally inhibited the wider uptake of 3DED. To address this limitation, we present novel *DigitalMicrograph* script *GiveMeED* appropriate for controlled 3DED data acquisition. The collection of useable reflections beyond 0.8 Å makes 3DED crystallographic processing effectively routine, using standard software and workflows derived from single-crystal X-ray diffraction (SCXRD) techniques. A full experimental workflow for 3DED on a conventional TEM is described in practical terms, in combination with direct imaging, and energy dispersive X-ray spectroscopy (EDS) and electron energy loss spectroscopy (EELS), for the return of comprehensive correlative descriptions of crystal morphologies and sample compositions, with due regard for the quantification of electron flux at each stage of the characterisation process. The accuracy and effectiveness of *GiveMeED* is demonstrated through structure solutions for case study paracetamol, copper(II) phthalocyanine, and percholorocoronene samples, characterised in their near-native states under controlled low dose conditions at either room or cryogenic temperatures, with determined unit cell parameters and atomic connectivity matching accepted literature X-ray structures for these compounds. To promote the wider adoption of 3DED, we make *GiveMeED* freely available for use and modification, in support of greater uptake and utilisation of structure solution procedures via electron diffraction.

## 1. Introduction

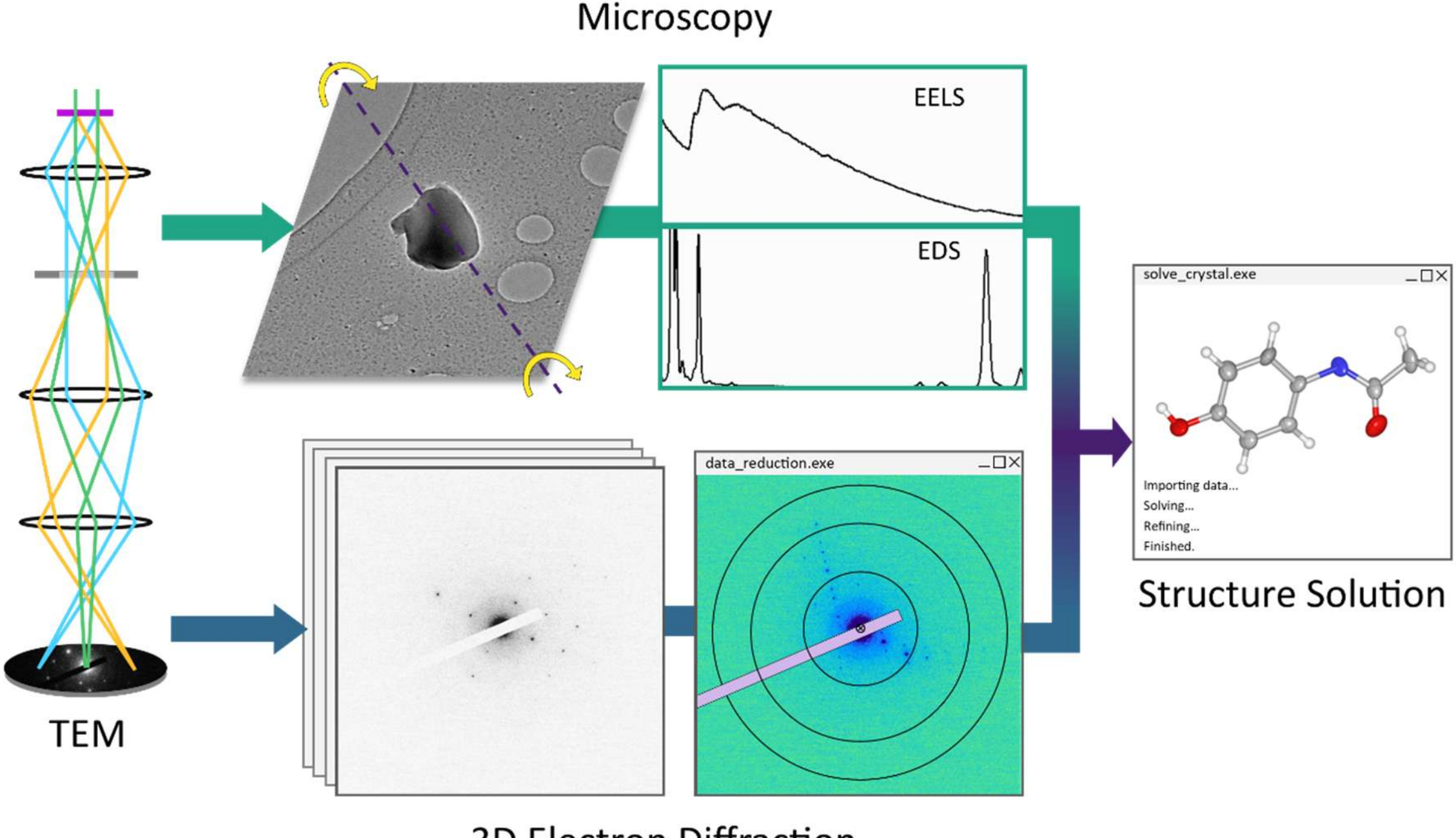

**Figure 1:** Illustration of the three-dimensional electron diffraction (3DED) workflow on a transmission electron microscope (TEM) used to solve crystal structures of small molecules. Diffraction patterns recorded using *GiveMeED* whilst a specimen is rotated are passed to downstream crystallography software and used to solve the structure of the compound. Complementary microscopy data (images and spectra), used to inform the solution/refinement process, also provide a wealth of additional information about the sample. The recording of high quality 3DED data is rendered facile, with standard operation protocols developed from practice and fostered by open knowledge outlined in Supplementary Information (SI).

Three-dimensional electron diffraction (3DED) provides for the analysis of nano- and microcrystals that are too small for single-crystal X-ray diffraction (SCXRD) or powder X-ray diffraction (PXRD) techniques (Figure 1). 3DED has been used successfully for structure determination of small molecules,[1–4] zeolites,[5–7] proteins,[8–15] coordination polymers,[16–18] coordination complexes,[19,20] and other compounds.[21–23] In addition to routine structure determination, 3DED has also been shown to be capable of advanced crystallographic analysis such as refining hydrogen positions.[24]

3DED methodology has undergone several iterations.[25–32] Rotation electron diffraction (RED)[33] utilised stepped rotation of both the microscope stage and electron beam to collect diffraction tomography series. Precession-assisted electron diffraction tomography (PEDT)[34] coupled stepped stage rotation with beam precession to give recorded intensities more similar to pure kinematic scattering intensities and increased crystallographic resolution, whilst the later innovation of continuous rotation electron diffraction (CRED) adopted continuous stage rotation to minimise the time needed for data collection whilst maximising the volume of reciprocal space interogated.[35] Data reduction in 3DED is currently performed by a small number of programs derived from SCXRD software: *CrysAlisPro,*[2] *PETS2,*[36] *Scipion-ED,*[37] *MicroED suite,*[38] *DIALS,*[39] and *XDS*.[40]

3DED has been performed using conventional TEM[3–5,17,21,33,41–44] operating at a range of accelerating voltages, using both field-emission gun and thermionic sources, and typically without aberration correction. TEMs enabled with charged-couple device (CCD),[5,45] complementary metal-oxide-semiconductor (CMOS),[5] or direct-electron detecting[3] cameras have all been used for data collection, under conditions of low electron flux to help maintain samples in their near-native state. Accordingly, 3DED may be used to investigate delicate crystal structures before the continuous loss of diffracted beam intensity by electron beam damage compromises their value.[46,47] This approach in TEM offers the additional advantage, compared to commercial electron diffractometers,[1,2] of detailed morphological investigation via high-resolution imaging combined with elemental analysis from the same volume of material, delivered via energy-dispersive X-ray spectroscopy (EDS) or electron energy loss spectroscopy (EELS). It is recognised that a lack of generalised 3DED setups means implementation requires competence in both crystallography and electron microscopy,[48] rendering it "…a highly specialised technique for microscopy experts".[1] This reflects the observation that competence in microscope operations and structure determination require separate skillsets, and the assumption that 3DED requires an individual to possess both. This reservation, however, may be overcome by the creation of more accessible methods for performing 3DED, and by establishing strong bridges between research communities, allowing crystal growers and crystallographers to make their data requirements clear and microscopists to design and carefully perform 3DED experiments. Crucially, collaboration allows microscopists and crystallographers to use their distinct skillsets synergistically.

Outside of some commercial solutions, 3DED is not typically performed using an integrated processing pipeline. Seamless integration of data collection with data processing requires commitment of skilled personnel and resources to develop and maintain both software and hardware, and its existence must be justified by a strong and continuous need to perform 3DED. By contrast, implementing 3DED via a unified workflow, with freely available software on local hardware, has a lower access cost that validates even sporadic demand for 3DED.

The vital missing piece to turn a TEM into an electron diffractometer is software to synchronise stage tilt and ED data acquisition. Compared to manual operation, automating data collection eliminates sources of user-error and provides for a consistent experimental protocol. Some 3DED software is commercially available[49] or open-source, *e.g. Instamatic*,[50] *InsteaDMatic*,[5] *Fast-ADT,*[43] and *SerialEM*,[51] but may not be applicable to local hardware/software setups. Hence, it's important that ED data collection software is easy to modify, so that it can be adapted and updated to future proof local instruments.

Here, we present a novel *DigitalMicrograph* script *GiveMeED* for recording continuous rotation 3DED datasets, with general applicability illustrated via three case study small molecule crystal structure solutions. For microscopists, we describe how to adapt TEMs to collect high quality diffraction data under conditions of low electron fluence; for crystallographers, three crystal structures solved using data collected via *GiveMeED* are described, demonstrating the effectiveness of 3DED compared to X-ray diffraction approaches; and for the broader community of researchers growing crystals, we emphasise that 3DED on a conventional TEM can be used effectively to provide crystal structure solutions for sub-micrometre sized samples unsuitable for X-ray investigation.

## 2. Materials and Methods

### 2.1. 3DED and TEM Data Acquisition Conditions

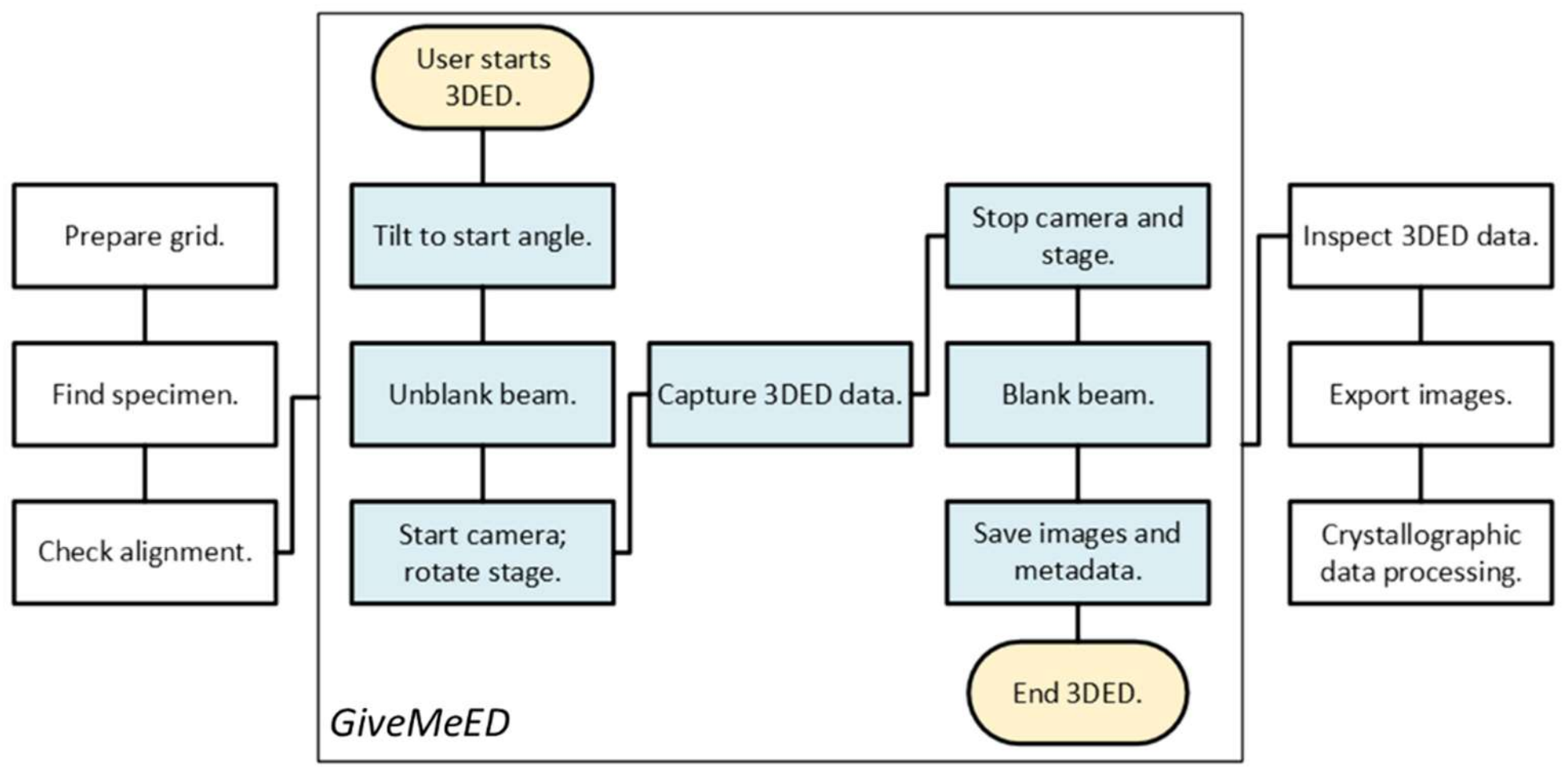

**Flow 1:** The workflow of a 3DED experiment. The steps executed by *GiveMeED* during 3DED data collection are shown in coloured boxes. Once the user presses “start”, the processes in blue boxes are automated – the microscope stage is rotated whilst the camera records diffraction data, which is then saved alongside necessary metadata. Automation gave a streamlined, repeatable data collection procedure, helping to reduce operator error.

3DED was performed using *GiveMeED* on a JEOL 2100Plus at 200 kV with a Gatan OneView camera at 175 fps, with image size 1024x1024. The camera was operated in both image (“I”) and video (“In-Situ”) modes for 3DED. Diffraction patterns were captured at 600 mm camera length (pixel size = 0.029 $nm^{-1}$) using selected area aperture 1 (100 µm diameter). The tilt-axis of the stage was calibrated prior to recording 3DED,[44] and could be recalibrated per-dataset; with the stage rotated at eucentric height using the lowest speed setting of around 9.1° $s^{-1}$. Specimens were located by low flux bright-field TEM imaging with defocussed objective lens (approx. -100 nm).

The pre-sample optics and emission current were used to control electron flux at the sample and the size of the illuminated area: *i.e.* brightness (C3 lens strength), alpha (convergence angle via condenser mini lens strength), spot size (minimum beam radius via condenser system), and emission current (“beam target”). Searching for samples was performed with 50 µm diameter condenser aperture, spot size 5, largest convergence angle, fully spread beam, and 104 µA emission current. Once a suitable sample was identified, the beam blank was used to minimise total electron fluence. 3DED was then performed with 100 µm diameter selected area aperture, spot size 1 or 2, and 108 µA emission current. The corresponding electron flux measured at the sample during 3DED is indicated per-specimen in Section 3.1. To minimise beam damage the electron flux while searching for samples was maintained around 0.01 $e^{-}$ $Å^{-2}$ $s^{-1}$. The electron beam stability of each sample was judged qualitatively by exposing a sacrificial region to an intense beam and observing how quickly the diffraction pattern faded. Where possible all alignments were performed without illuminating the target crystal. Limited beam exposure prior to 3DED occurred while finding the specimen, checking eucentric height, and confirming the specimen was crystalline.

During 3DED *GiveMeED* was used to control stage position, camera acquisition and metadata collection (Flow 1). Data was collected over the largest possible angle range at eucentric height, typically taking 12-14 seconds for 120° stage rotation. The stage was rotated in the same direction for each experiment. Mechanical backlash was accounted for by starting the camera acquisition after the stage had begun moving. To judge whether samples diffracted beyond 0.8 Å the script *AutoResRings* was used to draw resolution rings on the camera live view. Full details of all scripts are included in SI. To achieve high-quality crystal structures, data was recorded for samples that showed at least diffraction at 0.8 Å crystallographic resolution.

Complementary TEM images and associated spectra from the same sample were acquired following 3DED. Bright-field TEM imaging was performed with the OneView camera in both “I” and “Normal” modes, with 4K imaging performed at typically 1 s exposure and either 12 kX or 15 kX magnification (pixel size 0.024 nm or 0.77 nm). EELS was performed using a Gatan Quefina EL spectrometer, with typically 5 s acquisition time. EDS was performed with an Oxford Instruments X-Mantle detector typically with 1 minute live time. Samples were loaded at room temperature (RT) into the TEM column. JEOL single-tilt or HTR holders were used for 3DED investigations performed at room RT.  A Gatan Elsa cryo-tomography holder was used for cryoTEM investigations, with samples cooled to 120 K *in situ*.

**2.2. TEM Data Processing**

*DigitalMicrograph* (ver. 3.60.4435.0) was used for processing 3DED datasets into a format that could be imported into *PETS2* (ver. 2.2.20231114.1332), *CrysAlisPro* (ver. 171.43) and *DIALS* (ver. 3.23). The EDS spectra and images presented were prepared with *Digital Micrograph* (ver. 3.60). EELS spectra were background subtracted using a first order power law and plotted with *MATLAB* R2021a via the literature method.[52] Crystal structure models were prepared using *Vesta* (ver. 3.5.8).[53]

### 2.3. Data Reduction and Structure Solution/Refinement

Crystallographic processing of 3DED datasets was achieved with routine procedures and standard SCXRD software. Full details are given in the crystallographic information files and the SI (Table S1). Data reduction was performed with *CrysAlisPro* using a 0.6 Å resolution limit for integration and 0.8 Å for data finalisation, applying empirical absorption correction via *SCALE3 ABSPACK* (1.0.7). The output from *CrysAlisPro* was used for kinematic structure solution and refinement with *SHELXT* and *SHELXL* in *Olex2* (ver. 1.5)[54–56] Hydrogen atoms were geometrically placed and refined with a riding model, and rigid bond and similarity restraints were applied to the anisotropic displacement parameters of all structures. Geometric similarity restraints were applied to symmetry related bond distances such as phenyl ring moieties, and the anisotropic displacement parameters of some individual atoms were restrained to have more isotropic character. The solved structures were checked for missing symmetry using *ADDSYM* in *Platon*. Comparison to the literature X-ray structures was achieved through overlaying their asymmetric units and calculating the root mean square distance and the R-factors of the refined structures (Figure S1, Table S2).

### 2.4. Electron Flux Measurements

The camera gain was calibrated by measuring beam diameter for a range of beam currents, then plotting intensity versus flux and taking the gradient (Figure S2). Beam intensity was measured with the Gatan OneView camera; beam currents were measured at the maximum current value with a Keithley 6485 Picoammeter connected to a Faraday cup integrated into a Gatan Model 646 Double Tilt Analytical Holder. Electron flux measurements were performed according to the method in Fung *et al*.[57,58] An image was taken with a 1 s exposure, the mean intensity per pixel was measured for a region of vacuum, then converted to electron flux using the calibrated camera gain (Table S3). From this the applied fluence for each stage of the 3DED analytical procedure described, and the cumulative fluence over the entire analysis, was calculated for each sample (Table S4). Beam exposure time was measured using how long a detector was running, so fluence applied during scouting for samples and aligning the microscope for different analytical techniques was calculated using an estimated beam exposure time. For EDS and EELS, an estimated electron flux based on typical illumination conditions was used.

### 2.6. Sample Preparation

Methods of sample preparation have been discussed in literature.[48] Perchlorocoronene and copper(II) phthalocyanine were supplied pre-prepared on lacy carbon on copper 200 mesh TEM grids. Paracetamol was recrystallised from hot water prior to use, then gently ground and drop-cast from n-hexane onto a lacy carbon on copper 200 mesh grid.

## 3. Results

### 3.1. Solving crystal structures using *GiveMeED*

The crystal structures of paracetamol, perchlorocoronene (PCC) and copper(II) phthalocyanine (CuPC) were determined from data collected with *GiveMeED*. Complementary bright-field TEM images and EDS/EELS spectra were also recorded for all samples (Figures 2 – 4).

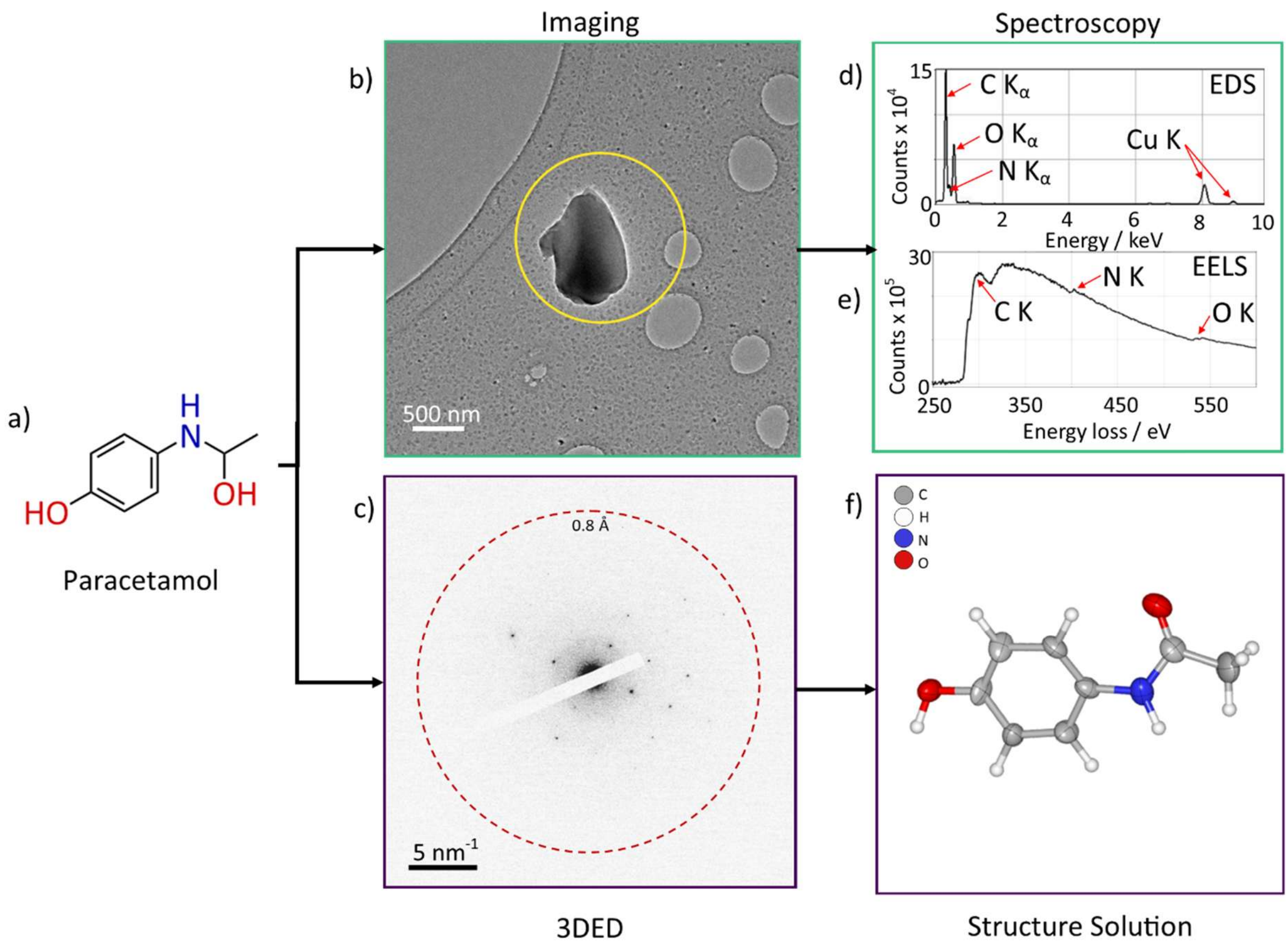

**Figure 2:** Graphical overview of how the structure of (a) paracetamol was solved via 3DED. (b) Low dose, cryogenic, bright-field TEM imaging located a suitable crystal (yellow circle indicates diffraction aperture position) then (c) 3DED data was recorded using *GiveMeED*. The dashed red circle indicates reflections beyond 0.8 Å resolution could be captured during 3DED. (d) EDS and (e) EELS confirmed the specimen contained nitrogen, oxygen and carbon, before (f) the structure was solved and refined from electron diffraction data.

The structure of paracetamol was solved from a single crystal grain at 120 K (Figure 2). The specimen was irregular, measuring 1.1 µm by 0.89 µm in projection. The target crystal was rotated 121° over 13.3 s with a mean electron flux of 1.90 e- $Å^{-2}$ $s^{-1}$ for a total electron fluence of 24.9 e- $Å^{-2}$. 1152 unique reflections were collected for 75.8 % completeness to 0.8 Å. The lattice parameters were determined to be: $a$ = 7.06 Å, $b$ = 9.19 Å, $c$ = 11.47 Å, α = 90 °, β = 96.2 °, γ = 90 °, with space group $P2_1/c$ (#14). For the anisotropic structure, $R_1$ was 23.8 % and $R_{int}$ was 13.3 %.

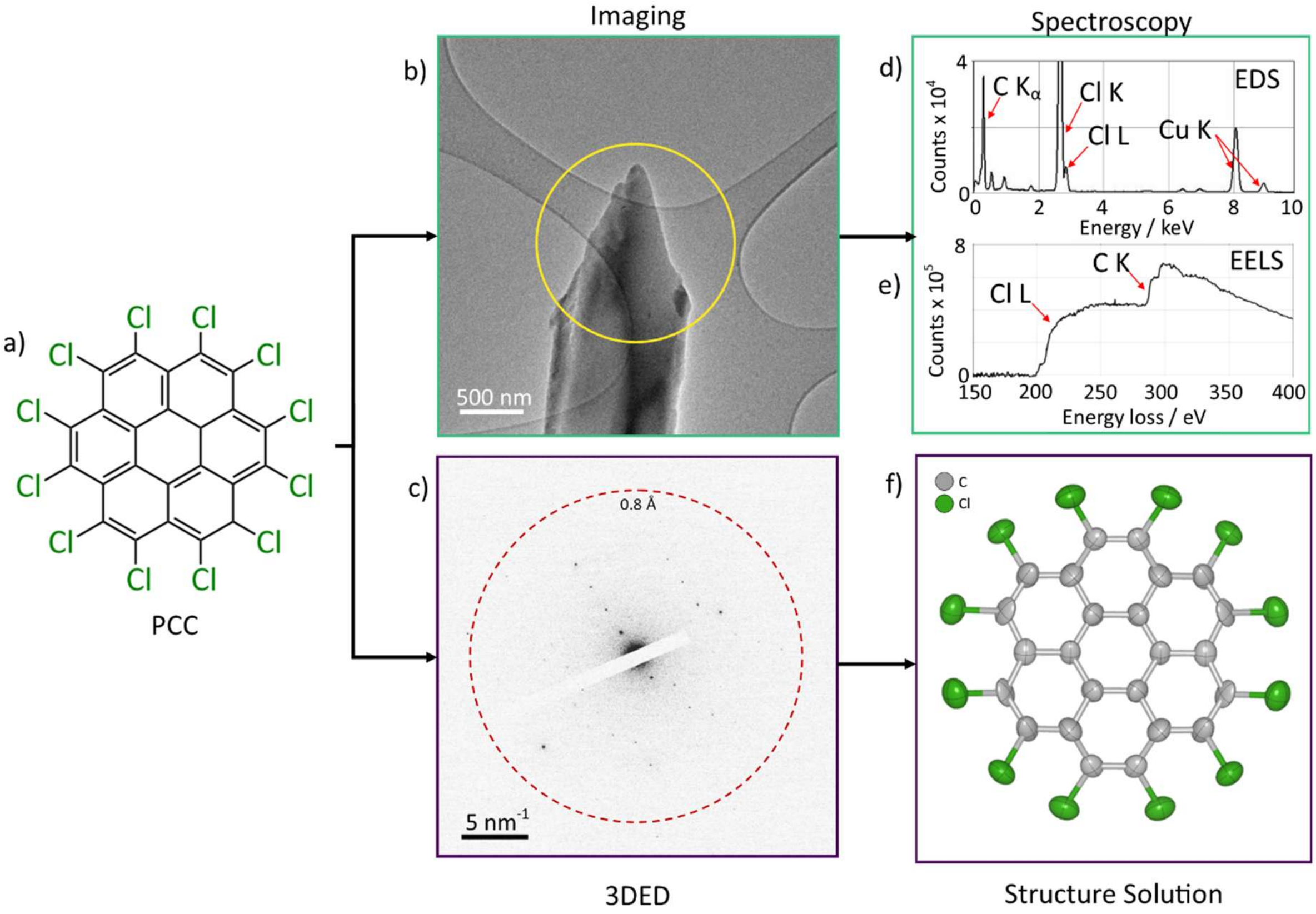

**Figure 3:** Graphical overview of how the structure of (a) perchlorocoronene (PCC) was solved via 3DED. (b) Target crystal found in bright-field TEM mode at cryogenic temperatures

(yellow circle indicates diffraction aperture), then 3DED data was acquired using *GiveMeED*. Dashed red circle indicates 0.8 Å resolution. (d) EDS and (e) EELS confirmed the presence of chlorine and $sp^2$ carbon in the sample prior to (f) structure solution and refinement from electron diffraction data.

The structure of PCC was solved from a single crystal grain at 120 K rotated 121° over 13.2 s (Figure 3). The specimen comprised a crystal needle, the tip of which was 0.98 µm at its widest point and 1.0 µm long. Diffraction was collected from the tip of the needle (Figure 3b). The mean electron flux was 1.69 e- $Å^{-2}$ $s^{-1}$ and total electron fluence was 22.3 e- $Å^{-2}$. 1213 unique reflections were collected for 99.0 % completeness to 0.8 Å. The lattice parameters determined were: $a$ = 22.37 Å, $b$ = 8.16 Å, $c$ = 12.81 Å, α = β = γ = 90 °, with space group *Cmce* (#64). $R_1$ for the refinement 33.3 % and $R_{int}$ was 13.1 % for the anisotropic structure.

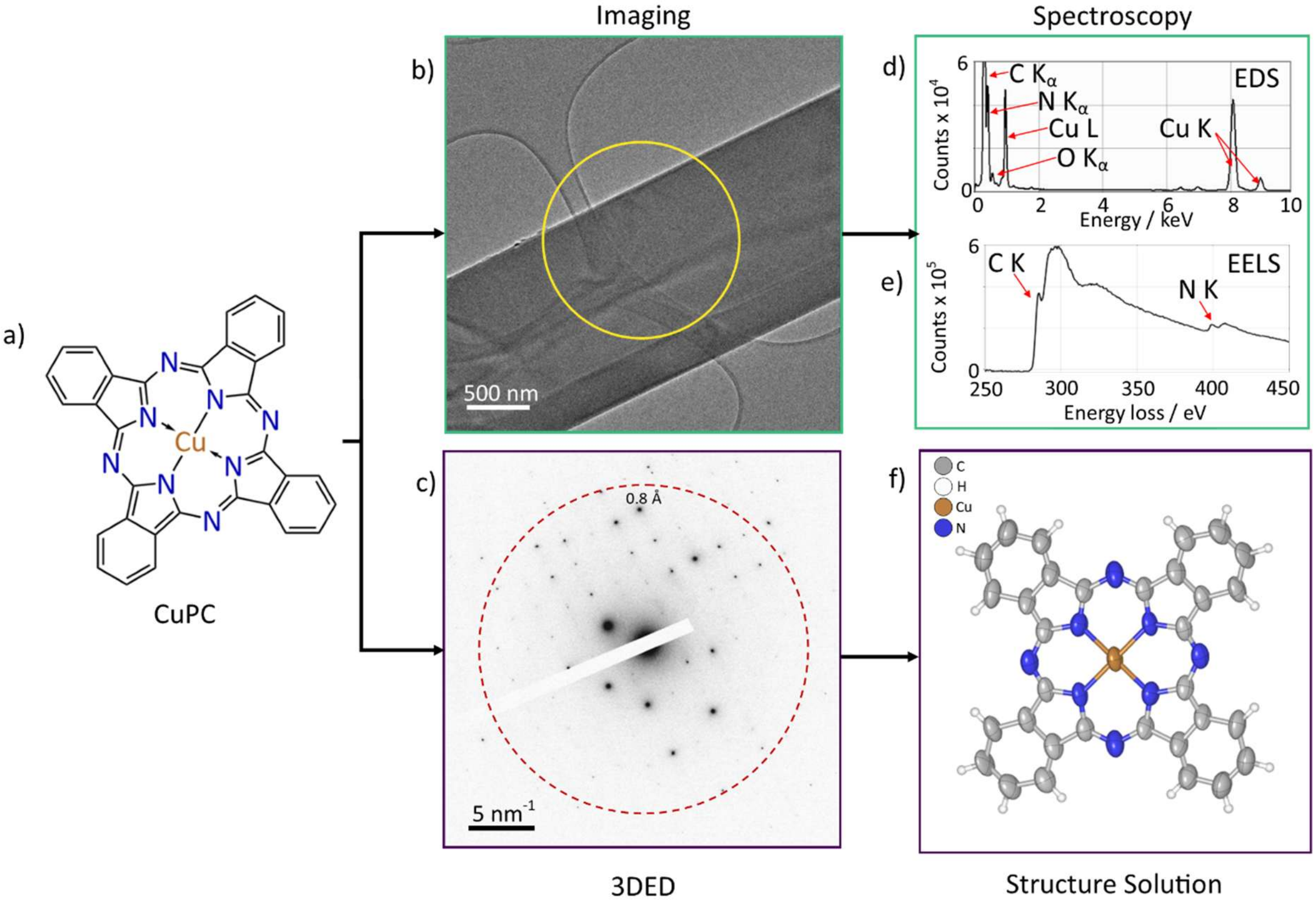

**Figure 4:** Structure solution for (a) copper(II) phthalocyanine (CuPC) from 3DED data. (b) A crystal was located at room temperature via bright-field TEM (yellow circle indicates diffraction aperture); and 3DED data was collected using *GiveMeED*. Dashed red circle

indicates 0.8 Å resolution. (d) EDS and (e) EELS confirmed the presence of copper, nitrogen, and $sp^2$ carbon prior to structure solution. (f) The structure was solved and refined from electron diffraction data.

CuPC was solved from three crystal grains at room temperature (Figure 4). The crystals were approximately 1.5 µm in width and several micrometres long, from which a small region was defined by the SAED aperture. 2137 unique reflections were collected for 94.5 % completeness to 0.8 Å resolution. Over all three grains, the mean flux was 0.62 e- $Å^{-2}$ $s^{-1}$ and the total electron fluence was 7.68 e- $Å^{-2}$. The lattice parameters determined were: $a$ = 14.29 Å, $b$ = 4.75 Å, $c$ = 17.0 Å, α = 90 °, β = 104.8 °, γ = 90 °, with space group $P2_1/c$ (#14). For the anisotropic structure, $R_1$ was 33.4 % and $R_{int}$ was 12.0 %.

## 4. Discussion

### 4.1 Overview

A novel script *GiveMeED* has been developed to collect 3DED datasets from sub-micron sized crystalline specimens within a conventional TEM, under low dose conditions at room or cryogenic temperatures, as appropriate, for the purpose of structure solution and refinement of samples maintained near their native state. In particular, 3DED combined with high spatial resolution TEM imaging and chemical analysis provided for the comprehensive investigation of individual specimens that would otherwise be intractable to identify and characterise using light microscopy and X-ray diffraction techniques. Three case study crystal structures, in the form of paracetamol, perchlorocoronene (PCC) and copper(II) phthalocyanine (CuPC) have been determined here, by way of illustration of the general applicability of 3DED.

In each case, initial low dose, bright-field TEM imaging during the scoping stage provided for quick assessment of sample crystallinity, via the presence of diffraction contrast observed with the objective lens defocused. Thereafter, the sequential workflow of 3DED data acquisition, mediated by *GiveMeED,* to provide datasets for the determination of sample structure, followed by correlative imaging and spectroscopy to confirm each specimen contained the elements expected, ensured time examining spurious contaminants was avoided, in advance of the more labour-intensive stage of crystallographic data analysis.

Data reduction, *i.e.,* the conversion of experimental diffraction patterns into a matrix of indexed reflections with measured intensity, required the development of protocols to convert 3DED datasets into a format that could be imported into standard X-ray software. Standard data reduction and structure solution/refinement procedures could then be used to solve the crystal structures from 3DED data; *i.e.,* the creation of atomic models describing experimental electron potential in the unit cell. Comparison with literature X-ray data[59–61] confirmed the structures returned by 3DED to be reliable, demonstrating the suitability of this workflow to solve molecular crystal structures, in particular crystals that cannot be grown with sufficient size suitable for characterisation via X-ray techniques.

It is recognised that both microscopy and crystallography are rich in technical terminology, hence plain language clarifications have been added to the text along with a list of abbreviations included in Supporting Information (SI Table S6).

### 4.2. ***GiveMeED***

*GiveMeED* is a 3DED data collection script for continuous-rotation electron diffraction measurements. Written in *DigitalMicrograph* scripting language, it is currently compatible with electron cameras controlled by *DigitalMicrograph* which support "In-Situ" imaging mode, and works by syncing the camera and stage rotation as the specimen stage is rotated through a user-defined angle range (Flow 1).

*GiveMeED* enables those comfortable with microscope operations to perform repeatable data collection using a consistent protocol, eliminating operator error and lowering the experience threshold required to perform 3DED. It also provides consistent methodology between different operators. For routine operation, a user interface has been provided. It was designed to be flexible via user modification, and adaptable to future developments in electron crystallography methodology. Hence, the code for *GiveMeED* is freely available and full operational details are included in SI.

*GiveMeED* has been designed to be understood easily and modified. Extending *GiveMeED* requires knowledge of the *DigitialMicrograph* scripting language, a high-level object-orientated interpreted language, which is expected to be straightforward for researchers familiar with languages such as Python or C++. Each process in the script has been compartmentalised into a function, letting users more easily adjust individual aspects of the data collection software, and to allow quick transplantation of functionality to new software. These design choices are intended to make development of new software based on *GiveMeED* relatively straightforward, alleviating some of the time and skill threshold required to enact automation.

*GiveMeED* records 3DED metadata in CIF format at the point of data capture. This contains information including electron wavelength and the rotational relationship between each diffraction pattern; *i.e.,* all the information content required to perform crystallographic analysis. Pathways have been developed to export data from *GiveMeED* into common data reduction software (*DIALS*, *PETS2* and *CrysAlisPro*) facilitated by scripts written in *DM*-script and Python. Flexibility is key, hence data recorded with *GiveMeED* can be converted into arbitrary image file formats, suitable for the individual needs of users and their preferred data reduction software. Such design choices allow *GiveMeED* to be modified as standard practices for electron diffractometry develop, empowering users to collect the exact data considered necessary for their scientific endeavours.

**4.3. Data collection workflow**

Accurate crystal structures required collection of high angle reflections from crystal planes with small d-spacing, giving sub-Ångstrom precision of atomic positions in the solved crystal structures. We aimed to collect reflections of at least 0.8 Å crystallographic resolution, judged using resolution rings on the camera live view, and routinely collected reflections corresponding to smaller d-spacing. This gave confidence that the data collected would be useable for solving crystal structures.

The microscope was operated under conditions of very low electron flux to limit beam damage whilst searching for samples, which meant defocus of the lens was required to ensure enough contrast, via Fresnel effects, to see the sample on the camera live view.

Once a suitable specimen was located, recording data was as simple as pressing "Start 3DED" on the *GiveMeED* user interface (Flow 1). Thereafter, qualitative TEM spectroscopy confirmed the presence of elements anticipated within the sample (Figure S3, Table S5), sufficient to support atom assignments performed during structure solution and refinement, without need for quantitative determination of elemental concentrations. This approach also provided for the filtering of datasets based on spectroscopy, before performing any labour-intensive crystallographic data analysis. Indexing a diffraction pattern will precisely identify a material but requires extraction of unit-cell parameters to match against a database of known structures. Spectroscopy, by contrast, can quickly eliminate an unsuitable sample, *e.g.* contamination, based on elemental composition. Overall, data collection became routine and the most time-consuming step was locating suitable crystals in each sample.

### 4.4. Electron fluence and beam damage

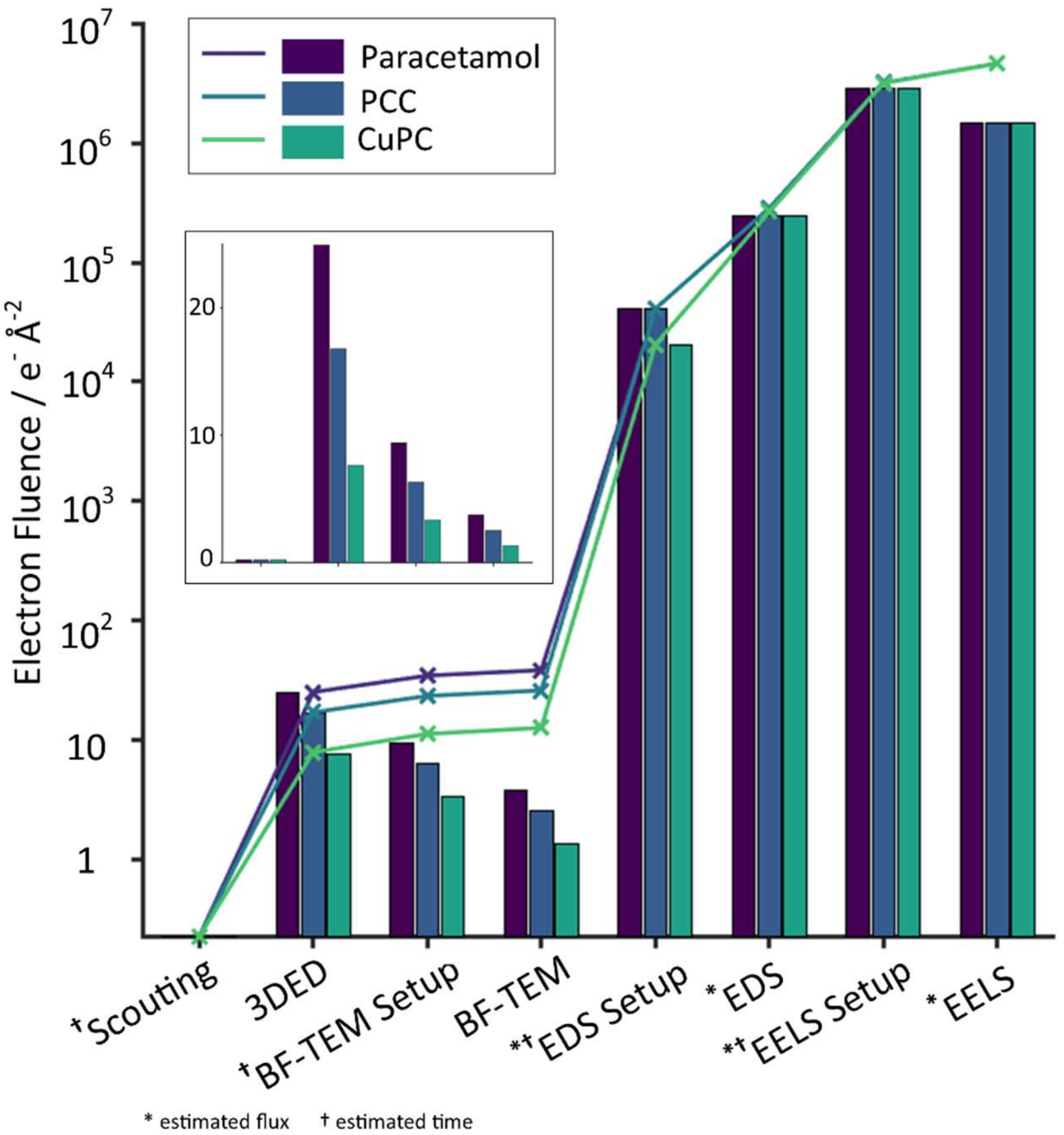

**Figure 5:** Electron fluence applied at each step of analysis (bar chart) and the cumulative electron fluence at each stage (line chart) for paracetamol (purple), PCC (blue), and CuPC (green). Inset shows the first four steps in greater detail. The majority of electron fluence was applied during spectroscopy, while 3DED and imaging involved a much smaller amount of fluence. Between each technique, "Setup" indicates the estimated fluence applied while the microscope was adjusted for a new mode of analysis. Fields marked with asterisk (*) were used estimated electron flux, and those with dagger (†) used estimated time, to calculate electron fluence.

In the context of electron diffraction, beam damage causes reflections to be lost from the diffraction pattern as the crystal lattice becomes amorphous - starting with high-angle reflections - often following an exponential decay in spot intensity proportional to the applied electron fluence. Hence, prior to and during 3DED measurements, minimum electron flux was used to preserve the crystal lattice at near native-state.

A thorough discussion of electron beam damage mechanisms and how to mitigate their effects has been given by Egerton.[62–65] Briefly, specific beam damage pathways are strongly influenced by the physical properties of the specimen. Especially for non-conductive samples, thicker specimens may be more susceptible to the effects of ionisation, secondary electrons, and heating. For atomically thin, conductive specimens the main source of damage is often direct knock-on damage leading to atom sputtering.

Mitigation against the effects of beam damage is vital for accurately recording intensities in diffraction patterns. Cryogenic temperatures have been shown to stabilise organic crystals by reducing the rate of structural rearrangement following damage. Reducing accelerating voltage typically reduces the rate of damage by changing the interaction cross-section, although increases in multiple electron scattering may cause reflection intensities to differ from pure kinematic scattering. Measured damage is typically linked to fluence but independent of flux, *i.e.* the same total fluence over a shorter time would cause the same measured amount of beam damage. In general, limiting the electron flux and beam exposure time is often effective at reducing the rate of beam damage to the specimen, although some counter-examples have been reported. [66] Notwithstanding, it is functionally impossible to perform 3DED without any prior electron beam exposure to the sample, hence the importance of minimising and monitoring electron fluence, and by implication damage, to ensure the crystal remains close to its native state during data acquisition.

By measuring the electron flux and the beam exposure time, the applied electron fluence during 3DED, imaging and spectroscopy can be calculated. Noting the electron fluence applied while changing the microscope between analytical modes must also be estimated to allow calculation of cumulative electron fluence over the entire analysis (Figure 5; Table S4). Whilst some measure of beam exposure during setup was inevitable for each mode of analysis, *e.g.* to confirm the EDS detector registered sufficient counts per second to acquire representative spectra, careful monitoring ensured the minimised development of damage.

Tracking of which techniques applied most fluence to the specimen clearly showed the importance of performing 3DED in advance of BF-TEM imaging and/or spectroscopy, as each technique increased the cumulative fluence, with increased risk of damaging the crystal structure and taking the specimen further from its native state. The process of spectroscopy clearly applied the most fluence, with the potential to cause the most damage to the sample. This was borne out in the diffraction pattern; after 3DED and TEM imaging, reflections above 0.8 Å were preserved indicating the structure to be still useful for crystallographic analysis. The use of a high electron flux for EELS and EDS analysis to facilitate the rapid collection of spectra often resulted in complete loss of reflections in the diffraction pattern. Whilst possible to collect spectra with a lower flux, to help preserve the crystal structure, this would necessitate a proportionate increase in time to collect spectra with acceptable signal-to-noise, hence the approach taken to prioritise sample throughput. In general, any degradation of the crystal lattice as a result of spectroscopy did not compromise the qualitative elemental information content provided by EDS and EELS, in support of 3DED structure determination. EELS was associated with the largest exposure during setup, as it required condensing the electron beam to gain a measurable core-loss spectrum, and so was performed last. Overall, careful fluence monitoring and control provided for confidence in the quality of the crystallographic data acquired, by limiting the onset, development and extent of sample damage.

**4.5. Electron crystal structure determination**

The structures of three small molecules solved from sub-micron sized specimens, which would not have been possible to examine using SCXRD techniques. As the data collected contained useable reflections above 0.8 Å, crystallographic processing was routine, using standard software and workflows derived from SCXRD. Despite the known practical challenges of electron crystallography,[26] we determined unit cell parameters and atomic positions that matched the literature X-ray structures for the compounds in this study.[59–61] Hence, we were able to identify correctly the atomic connectivity of the materials.

The refinement indicators are typical for kinematically refined electron crystal structures. Treatment of dynamical diffraction intensities with a kinematic refinement model leads inherently to greater uncertainty in unit cell parameters and atomic positions, as the experimental diffraction intensities do not match well to the calculated kinematic intensities, seen particularly in the relatively high R-factors for each structure when compared to X-ray data (Table S2). This was partially compensated by applying an extinction correction. Further improvements to the structure statistics were achieved by applying rigid bond and similarity restraints to the anisotropic displacement parameters of refined structures, i.e. using chemical intuition to ensure the structures made physical sense. Dynamical refinement has been shown to give improved accuracy in atomic positions,[67] but was not necessary here as the kinematic refinement was sufficient to confirm the identity of the crystals.

Comparing the electron structures with the literature X-ray structures confirmed that the electron structures were reliable, with the evaluation performed using metrics the demonstrated by Klar *et al*.[67] Overall, the X-ray structures had lower R-factors indicating better agreement between model and experimental data, even though electron data for paracetamol and PCC were recorded at lower temperatures than the corresponding X-ray data. This highlights the complementary nature of X-ray and electron diffraction – while X-ray structures have better statistics, electron structures can be acquired for samples too small for X-ray analysis. Overlaying the asymmetric units of the electron and X-ray structures revealed good alignment between the two datasets (Figure S1); while X-ray data had more precise anisotropic displacement parameter, those from the electron structures map well to the X-ray data as seen by the root mean square distance values for the overlapping fragments (Table S2). None of these results are unexpected, as X-ray structures are solved and refined from kinematic diffraction intensities. Hence, we have confidence in the utility of 3DED data recorded with *GiveMeED* to generate useful crystal structures in future studies.

### 4.6. Comparison to other software

Direct comparison of 3DED data collection software, *e.g.* to establish software performance, is limited by a number of confounding variables. Metrics from crystal structure determination, such as R-factors, are influenced by the data processing steps; for example, Vypritskaia *et al.* [68] noted that detector gain settings used in data reduction impacted their ability to solve the structure of natrolite based on the version of *SHELXT* used. Further, metrics such as electron flux or stage rotation speed reflect the instrumentation used. Hence, only a holistic comparison of the entire 3DED process is possible, which may give misleading conclusions about the 3DED data collection step. Nevertheless, if the final crystal structures are chemically sensible with reasonable refinement statistics, one can be confident in the data collection.

## 5. Conclusions

3DED is primed to become a major tool for crystallographers and microscopists, for the determination of accurate crystal structures for compounds that cannot be investigated using X-ray techniques. The major roadblock to 3DED uptake is the availability of instrumentation - to this end the *DigitalMicrograph* script *GiveMeED* was developed to allow adaptation of conventional TEMs to perform electron diffractometry. A workflow for collecting 3DED data sets to 0.8 Å resolution at low electron flux has been described so that electron microscopists unfamiliar with crystallography can collect high-quality data using *GiveMeED*. For crystallographers, the structures of paracetamol, perchlorocoronene, and copper(II) phthalocyanine were presented, solved from electron diffraction data, demonstrating that crystal structures matching to literature X-ray structures can be produced from data recorded with *GiveMeED* on a TEM. A full workflow from sample preparation to data analysis has also been described, demonstrating that elemental information from EDS and EELS can also be acquired from the same specimen volume in support of 3DED. We anticipate that 3DED will continue to develop, as more facilities acquire capacity to collect electron diffraction data for structure solution, with *GiveMeED* contributing as a foundation for 3DED data collection.

## Acknowledgements

*GiveMeED* is freely available online at https://github.com/benweare/GiveMeED. The structures of the compounds in this work are available from The Cambridge Crystallographic Data Centre (www.ccdc.cam.ac.uk/structures) with reference numbers 2464563, 24564, and 24565. Diffraction patterns used to solve crystal structures are available at https://doi.org/10.17639/nott.7567. The authors would like to thank the Nanoscale and Microscale Research Centre (nmRC) and staff for technical support and useful conversations; the School of Chemistry at the University of Nottingham and staff for technical support; Dr A. Nasir for useful conversations; and Dr J. P. Tidey for useful conversations.

## Author Contributions

**B. L. Weare:** Conceptualization, Methodology, Software, Validation, Formal analysis, Investigation, Data Curation, Writing – Original Draft, Visualisation, Project administration. **K. L. Y. Fung:** Visualisation, Writing – Review and Editing. **I. Cardillo-Zallo:** Investigation, Data Curation, Resources, Writing – Review and Editing. **W. J. Cull:** Visualisation, Writing – Review and Editing. **M. W. Fay:** Methodology, Software, Validation, Project administration. **S. P. Argent:** Supervision, Formal analysis, Validation, Writing – Review and Editing. **P. D. Brown:** Conceptualization, Methodology, Supervision, Project administration, Funding acquisition, Writing – Review and Editing.

**Funding Information**

Funding for this research was provided by: Engineering and Physical Sciences Research Council (grant No. EP/W006413/1 and EP/L022494/1); Leverhulme Trust (RPG-2022-300).

**Supporting Information for Practical Crystallography with a Transmission Electron Microscope**

Benjamin L Weare,[1,2] Kayleigh L Y Fung,[3*] Ian Cardillo-Zallo,[1,4] William J Cull,[2†‡] Michael W Fay,[1,5] Stephen P Argent,[2] Paul D Brown[5]

1 - Nanoscale and Microscale Research Centre, University of Nottingham, Nottingham NG7 2RD, United Kingdom.
2 - School of Chemistry, University of Nottingham, Nottingham NG7 2RD, United Kingdom.
3 - Department of Computer Science, Nottingham Trent University, Nottingham NG11 8NS, United Kingdom.
4 - School of Pharmacy, University of Nottingham, Nottingham NG7 2RD, United Kingdom.
5 - Department of Mechanical, Materials, & Manufacturing Engineering, Faculty of Engineering, University of Nottingham, Nottingham NG7 2RD, United Kingdom.

* Bruker UK, Electric Works, 3 Concourse Way, Sheffield, S1 2BJ, United Kingdom.
† SuperSTEM, SciTech Daresbury Science and Innovation Campus, Daresbury, WA4 4AD, United Kingdom.
‡ School of Physics, Engineering, and Technology, University of York, York, YO10 5DD, United Kingdom.

## 1. Overview

This Supporting Information contains in-depth instructions for using the script *GiveMeED* to collect 3DED data, and details of the scripts used to support 3DED data collection and analysis *AutoResRings*, *Go2Alpha,* and *ExportInSitu*. Details are given of how to import data recorded with *GiveMeED* into 3DED data processing software. A complete description of how to measure electron flux using a CMOS camera via determining it gain is also provided, applicable to any CMOS or CCD camera. Supporting EDS and EELS spectra are presented. *GiveMeED* has been demonstrated on a JEOL 2100Plus TEM with Gatan OneView camera, and JEOL 2100F TEM with Gatan K3-IS camera. All software is available via the associated GitHub: https://github.com/benweare/GiveMeED.

## 2. Crystallographic Data Processing

| | **Paracetamol** | **PCC** | **CuPC** |
|---|---|---|---|
| Formula | $C_8H_9NO_2$ | $C_{24}Cl_{12}$ | $C_{32}H_{16}CuN_8$ |
| Merged grains | 1 | 1 | 3 |
| Tilt range / ° | 120.9 | 120.9 | 121.0; 120.5; 96.7 |
| Angle per frame / ° | 0.052 | 0.052 | 0.055; 0.055; 0.057 |
| Frames | 2333 | 2328 | 2180; 2188; 1685 |
| Mean flux / e- $Å^{-2}$ $s^{-1}$ | 1.88 | 1.27 | 0.47; 0.62; 0.91 |
| Total time / s | 13.3 | 13.2 | 12.5; 12.5; 9.68 |
| Temp / K | 120 | 120 | Room temp. |
| Resolution cutoff / Å | 0.8 | 0.8 | 0.8 |
| Reflections | 2784 | 3091 | 7774 |
| Unique Reflections | 1147 | 1213 | 2122 |
| Unit cell | 7.06(1), 9.19 (13), 11.47(9), 96.2(2) | 22.37(16), 8.16(14), 12.81(15) | 14.29(11), 4.75(8), 17.0(1), 104.8(3) |
| Space group | $P2_1/n$ | Cmce | $P2_1/n$ |
| Completeness / % | 75.8 | 99.0 | 93.4 |
| I/sig(I) | 8.2 | 7.7 | 9.7 |
| $R_{int}$ / % | 13.3 | 13.1 | 12.0 |
| R1 / % | 23.8 | 33.2 | 33.4 |
| wR2 / % | 58.7 | 73.8 | 66.3 |
| GooF | 2.26 | 3.20 | 2.51 |

**Table S1:** Data collection and crystallographic details for the case study structures of paracetamol, percholorocoronene (PCC), and copper(II) phthalocyanine (CuPC) presented in this work.

| | **Paracetamol** | | **PCC** | | **CuPC** | |
|---|---|---|---|---|---|---|
| | **ED** | **SCXRD** | **ED** | **SCXRD** | **ED** | **SCXRD** |
| R1 / % | 23.8 | 5.5 | 33.2 | 3.26 | 33.4 | 3.3 |
| σ / e- $Å^{-1}$ | 0.055 | 0.043 | 0.105 | - | 0.105 | - |
| Temp. / K | 120 | 150 | 120 | 295 | RT | 103 |
| RMSD / Å | 0.357 | | 0.639 | | 0.190 | |

**Table S2:** Comparison of R-factors, difference electron density map standard uncertainty (σ),[1] root mean square distance (RMSD) for overlaid asymmetric units, and experimental temperature for electron and literature X-ray structures.[2–4] For the X-ray structures of PCC and CuPC, no reflections were included in crystallographic information files so it was not possible to calculate the difference maps.

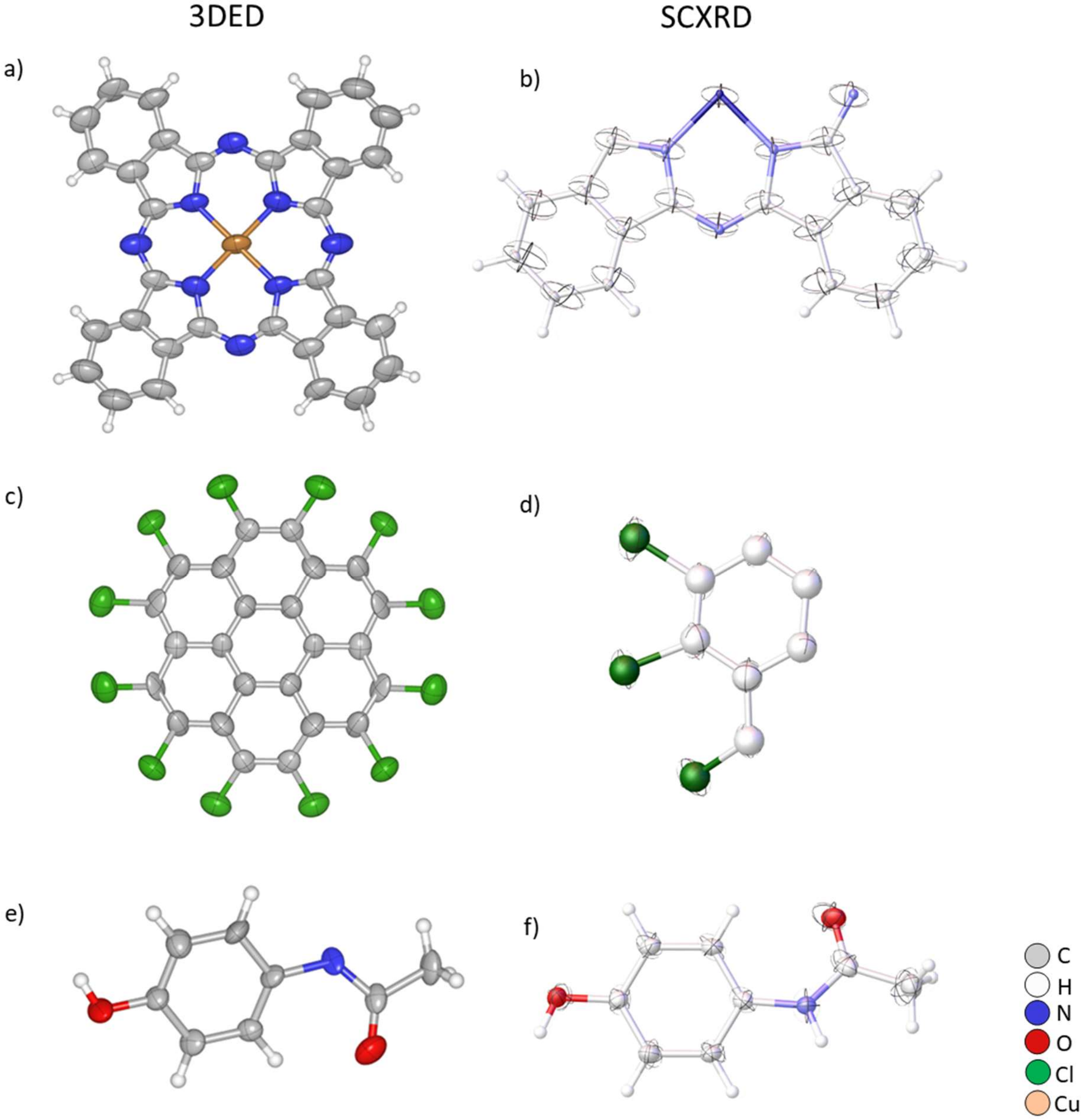

**Figure S1:** 3DED structures (left column) and overlay of 3DED and literature SCXRD asymmetric units (right column)[2–4] for (a, b) copper phthalocyanine, (c, d) percholorocoronene, and (e, f) paracetamol. It can be seen that the atomic positions from 3DED map to those from SCXRD, but with larger anisotropic displacement parameters arising from kinematic refinement of dynamical data.

Data reduction of 3DED datasets was performed with *CrysAlisPro*. Peak hunting used the following options: "3D peak extraction" with "Single frame background subtraction" and "Weak features extraction". A resolution limit of 0.6 Å was applied for integration. Data finalisation was performed under standard (kinematic) settings, with a resolution limit of 0.8 Å. Empirical absorption correction was applied using *SCALE3 ABSPACK* (1.0.7).

For all structures, kinematic solution and refinement was performed with *SHELXT* and *SHELXL* in *Olex2* (ver. 1.5) (Table S1).[5–7] Applied *SHELXL* instructions are indicated in parentheses. Hydrogen atoms were observed in the electron potential map then geometrically placed using neutron bond distances and refined with a riding model.[8] Rigid bond (*SHELXL* RIGU) and similarity (*SHELXL* SIMU) restraints were applied to the anisotropic displacement parameters of all atoms in the structures. The values of refinement statistics (R1, wR2, goodness-of-fit, extinction coefficient, and $R_{int}$) were high due to the structure being solved and refined from electron diffraction data using a kinematic model that did not account for the dynamical diffraction of electrons. This lead to greater uncertainty in atomic positions and unit cell parameters, *i.e.* structures confirmed atomic connectivity but were not suitable for extracting detailed geometric information due to the large uncertainties on bond lengths and unit cell parameters.

Experimental limitations of the electron diffraction data collection method (*i.e.* limited goniometer rotation range) meant the entire diffraction sphere was not sampled. Hence, a number of reflections were not present in the data used for structure solution and refinement. Some reflections had observed intensity ($I_{obs}$) much smaller than calculated intensity ($I_{calc}$) due to measurement discrepancies arising from the effect of dynamical diffraction or reflections impinging on the beam stop. *Platon ADDSYM* was used to test all structures for missing symmetry.[9]

For paracetamol, the structure was solved in space group $P2_1/m$ and reset in $P2_1/n$ using *Platon ADDSYM*. The anisotropic displacement parameter of amide carbon atom C8 was restrained to have more isotropic character (*SHELXL* ISOR). Geometric similarity restraints were applied symmetry related bond distances with the phenyl ring moiety (*SHELXL* SAME), where 1,2-bond distances used estimated standard deviation (e.s.d) of 0.02 Å and 1,3-bond distances used e.s.d of 0.04 Å.

For percholorcoronene, the anisotropic displacement parameter of amide carbon atom C1 was restrained to have more isotropic character (*SHELXL* ISOR). For copper(II) phthalocyanine, the structure was solved in space group Pm and reset in $P2_1/n$ using *Platon ADDSYM*. Data from three crystal grains was merged into a single dataset to increase data completeness using Proffit merge (*CrysAlisPro*). Same distance restraints (*SHELXL* SADI) were applied to chemically identical environments in the two benzene moieties of the asymmetric unit, where 1,2-bond distances used e.s.d of 0.02 Å and 1,3-bond distances used e.s.d of 0.04 Å.

### 3. Measuring Electron Flux and Calibrating Camera Gain

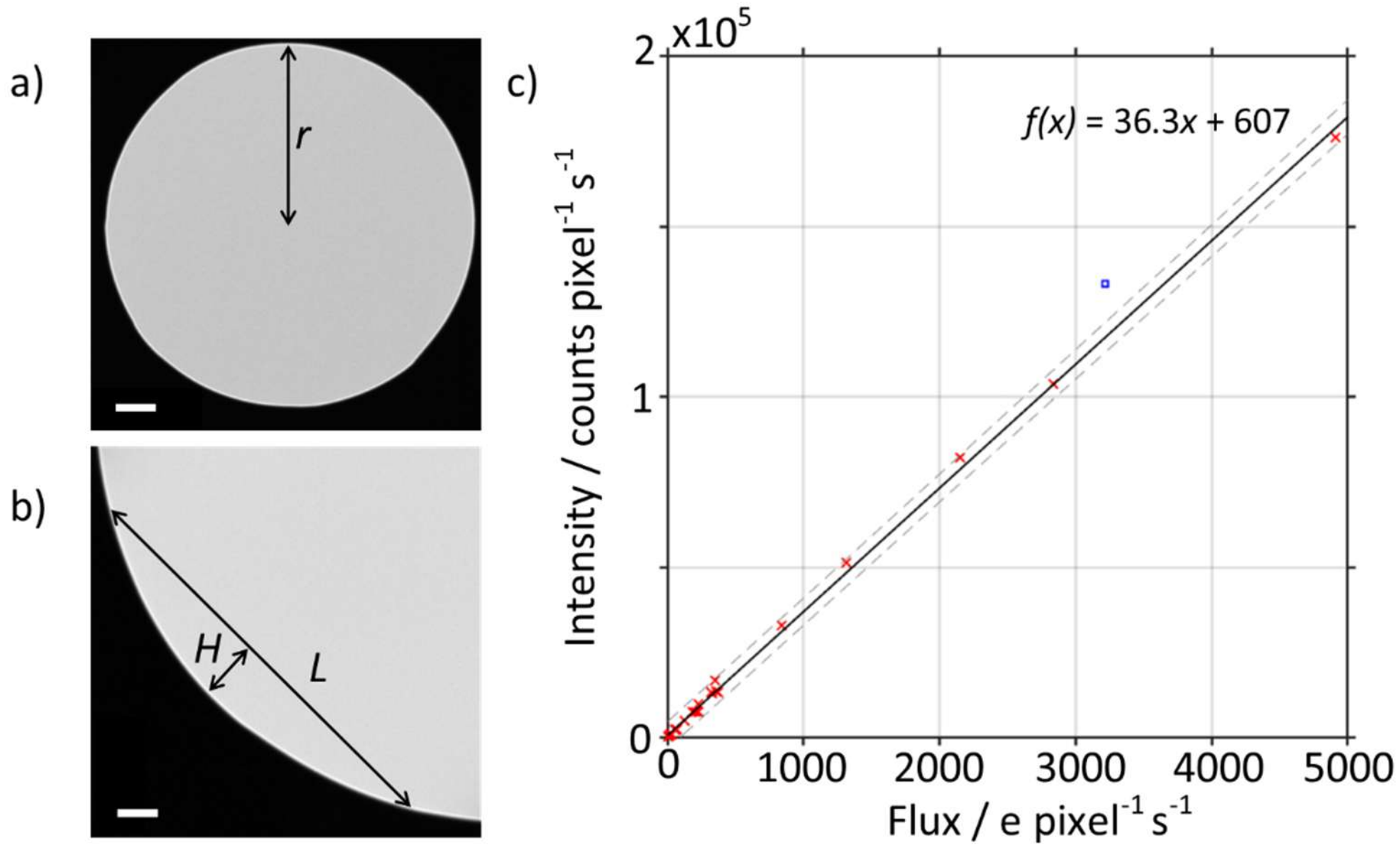

**Figure S2:** Electron micrographs of vacuum illustrating how electron beam diameter was measured using (a) the whole beam and (b) an arc of the beam via the intersection chords theorem (Equation S2). Scale bars are 500 nm. (c) Linear fitting of flux versus camera intensity used to measure camera gain. The gradient of the black fit line is camera gain ($G$) used to convert from intensity to flux: $G$ = 36.3 ± 0.8 counts electron$^{-1}$ (3σ standard deviation). Dashed grey lines indicate 99% confidence intervals. Red crosses are data points, and the blue square is an outlier ($R^2$ = 0.9987).

For accurate electron flux measurement on a CCD or CMOS cameras it was necessary to calibrate the conversion of camera counts to electrons (*i.e.* camera gain). Flux was measured using the method described by Egerton,[10] where flux $j$ is the ratio of the beam current $I_b$ and beam area $A_b$

$$j = \frac{I_b}{A_b} = \frac{I_b}{e\pi r^2} \tag{S1}$$

Where $r$ was the radius of the beam in pixels and $e$ was elementary charge, giving $j$ in electrons pixel$^{-1}$ s$^{-1}$. Electron beam radius was measured using the camera. When the entire beam fit on the camera sensor the beam radius was measured directly, taking the edge of the electron beam as the full width tenth maximum. To measure the radius of beams larger than the camera sensor, the intersecting chords theorem was applied

$$r = \frac{L^2}{8H} + \frac{H}{2} \tag{S2}$$

Where $L$ was chord length and $H$ was chord height (Figure S2a,b). This allowed measurement of large radius electron beams.

For each electron flux, the electron beam intensity at the camera ($Q$) was measured by taking the average intensity over a region of vacuum. A plot of $Q$ versus $j$ yielded the camera gain ($G$, in counts electron$^{-1}$) as the gradient (Figure S2c)

$$G = \frac{Q}{j} = \left(\frac{Q\pi r^2}{I_b}\right) \tag{S3}$$

At 200 kV, $G$ = 36.3 ± 0.8 counts electron$^{-1}$. To measure the electron flux in experimental data the electron beam intensity of vacuum was measured at a region of vacuum and applying the following relation

$$j_E = \frac{Q_E}{d^2 G} \tag{S4}$$

Where $j_E$ was the experimental electron flux, $Q_E$ was the experimental intensity, and $d$ was pixel size in nanometres per pixel. A look up table of $d^2G$ was then produced for a range of magnifications (Table S3). Experimental electron flux was used to calculate applied electron fluence ($F_E$) (Table S4)

$$F_E = j_e t \tag{S5}$$

Where $t$ was the length of time the specimen was exposed to flux $j_e$. The cumulative fluence ($F_C$) was then the sum of each electron fluence

$$F_C = \sum F_E \tag{S6}$$

| Magnification / kX | *d* / nm pixel$^{-1}$ | $d^2G$ / e nm$^2$ count$^{-1}$ pixel$^{-2}$ |
|---|---|---|
| 100 | 0.11 | 3.98 |
| 80 | 0.14 | 4.99 |
| 60 | 0.18 | 6.67 |
| 50 | 0.22 | 8.03 |
| 40 | 0.28 | 10.1 |
| 30 | 0.38 | 13.7 |
| 25 | 0.47 | 16.9 |
| 20 | 0.61 | 22.3 |
| 15 | 0.78 | 28.4 |
| 12 | 0.94 | 34.0 |
| 10 | 1.2 | 40.3 |

**Table S3:** Pixel length ($d$) and intensity-to-flux conversion factors ($d^2G$) for a range of magnifications at 200 kV. The conversion factors are the product of the gain ($G$ = 36.3 counts electron$^{-1}$) and the square of the pixel length.

| | Fluence / e- Å $^{-2}$ | | | Cumulative Fluence / e- Å $^{-2}$ | | |
|---|---|---|---|---|---|---|
| | **Paracetamol** | **PCC** | **CuPC** | **Paracetamol** | **PCC** | **CuPC** |
| †Scouting | $0.22 \pm 1.1$ | $0.22 \pm 1.1$ | $0.22 \pm 1.1$ | $0.22 \pm 1.1$ | $0.22 \pm 1.1$ | $0.22 \pm 1.1$ |
| 3DED | $25 \pm 0.076$ | $22 \pm 0.076$ | $7.6 \pm 0.088$ | $25 \pm 1.1$ | $22 \pm 1.1$ | $7.9 \pm 1.1$ |
| †Image setup | $9.4 \pm 0.2$ | $6.4 \pm 0.2$ | $3.4 \pm 0.2$ | $35 \pm 1.1$ | $23 \pm 1.1$ | $11 \pm 1.1$ |
| BF-TEM | $3.8 \pm 0.5$ | $2.5 \pm 0.5$ | $1.3 \pm 0.017$ | $39 \pm 1.2$ | $26 \pm 1.1$ | $12 \pm 1.2$ |
| *†EDS setup | $(4.1 \times 10^4) \pm 0.1$ | | | $(4.1 \times 10^4) \pm 1.2$ | $(4.1 \times 10^4) \pm 1.2$ | $(2.1 \times 10^4) \pm 1.3$ |
| *EDS | $(2.5 \times 10^5) \pm 0.017$ | | | $(2.9 \times 10^5) \pm 1.2$ | $(2.9 \times 10^5) \pm 1.2$ | $(2.7 \times 10^5) \pm 1.3$ |
| *†EELS setup | $(2.9 \times 10^6) \pm 0.1$ | | | $(3.2 \times 10^6) \pm 1.2$ | $(3.2 \times 10^6) \pm 1.2$ | $(3.2 \times 10^6) \pm 1.3$ |
| *EELS | $(1.5 \times 10^6) \pm 0.2$ | | | $(4.8 \times 10^6) \pm 1.3$ | $(4.8 \times 10^6) \pm 1.3$ | $(4.7 \times 10^6) \pm 1.3$ |

**Table S4:** Electron fluence for each step in the 3DED workflow, and the cumulative electron fluence at each step to two significant figures. Graphical representation of the tabulated data is included in the main text Figure 5. Errors were calculated according to Section 4. Rows marked with asterisk (*) used estimated electron flux, and rows with dagger (†) used estimated time, to calculate the electron fluence.

## 4. Calculation of Electron Fluence Errors

Calculation of errors in electron fluence used the following method. All standard deviations were to 99% confidence intervals unless otherwise stated, and error in time was taken as ± 1 second. The error in electron flux ($\varepsilon_j$) was calculated using the standard deviation of the measured pixel intensity ($\sigma_Q$) and the standard deviation of the camera gain ($\sigma_G$)

$$\varepsilon_j = \sqrt{\frac{{\sigma_G}^2}{G^2} + \frac{{\sigma_Q}^2}{Q^2}} \tag{S7}$$

The error in electron fluence ($\varepsilon_F$) was calculated from the error in electron flux and the error in time ($\varepsilon_t$)

$$\varepsilon_F = \sqrt{\frac{{\varepsilon_j}^2}{j^2} + \frac{{\varepsilon_t}^2}{t^2}} \tag{S8}$$

The error in cumulative electron fluence ($\varepsilon_{CF}$) was calculated from the sum of fluence errors

$$\varepsilon_{CF} = \sqrt{\sum {\varepsilon_F}^2} \tag{S9}$$

## 5. Exporting 3DED data from Digital Micrograph

3DED was recorded as .dm4 image files in the "In-Situ" dataset format. 3DED data was exported using the *DigitaMicrograph* script *ExportInSitu* and the *export_insitu* Python package; or manually, using the "In-Situ Editor" in *DigitalMicrograph* using the "Files to Folder" option to save each frame in the dataset as an individual 4-byte real image. A beam stop mask was added after images were imported into data reduction software.

To import data into *DIALS*, the images were saved as a single .dm4 stack. Experimental parameters were passed to *DIALS* via the command line as discussed in the literature; data reduction was then possible following the literature procedure.[11]

To import data into *PETS2* the "In-Situ" datasets were first exported in .tif format. A .pts2 project file for the dataset was created following the guidelines in the *PETS2* documentation; data reduction was then completed as per the literature procedure.[12]

To import data into *CrysAlisPro*, "In-Situ" datasets were exported in .dat format using the "Esperanto Importer" feature. The image size, byte type, rotation axis, and electron wavelength were not read from image headers so were supplied manually.

## 6. Energy Dispersive X-ray and Electron Energy Loss Spectroscopy

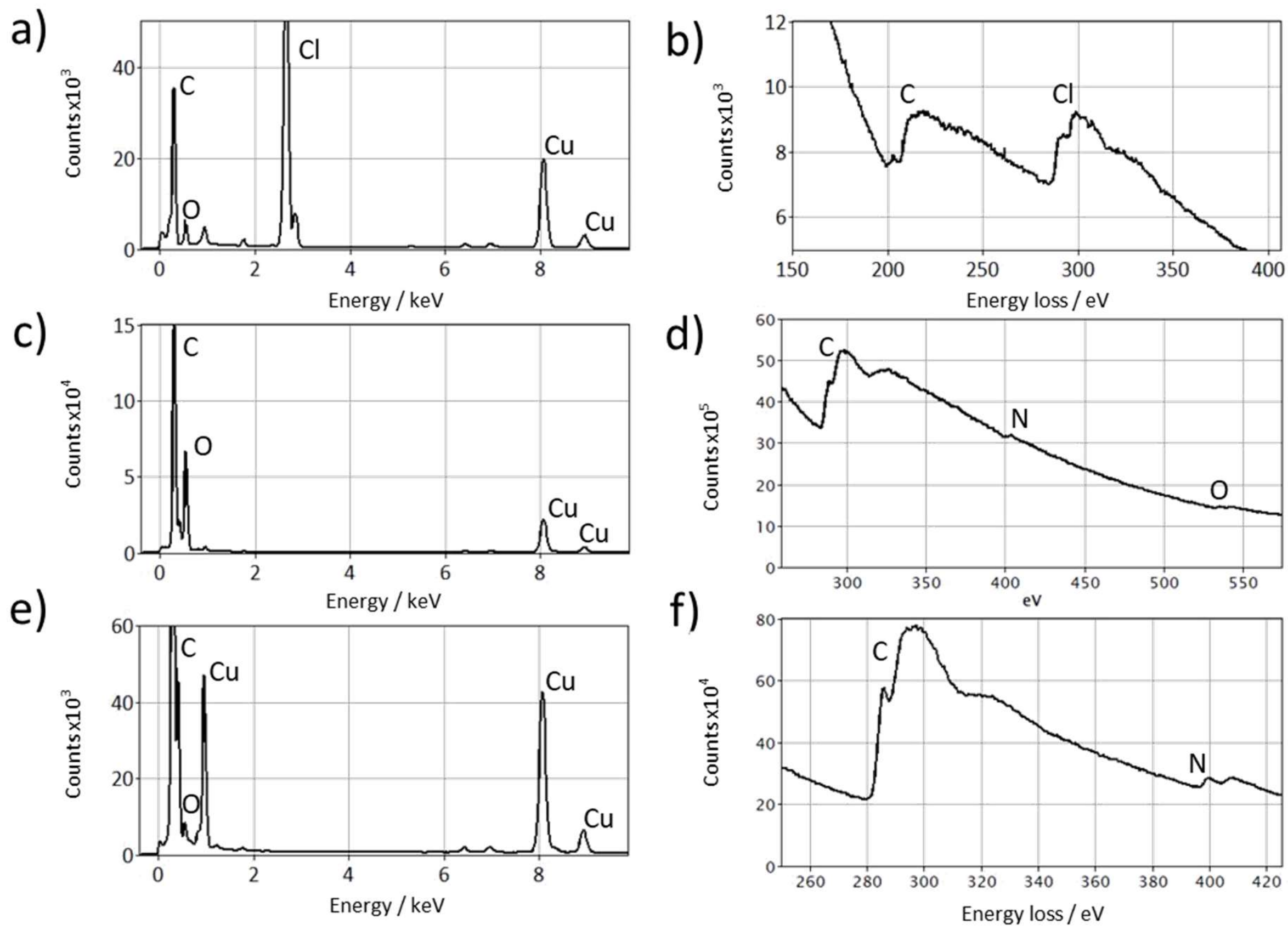

**Figure S3:** Energy dispersive X-ray (EDS, left) and electron energy core-loss (EELS, right) spectra for (a, b) percholorcoronene, (c, d) paracetamol, and (e, f) copper(II) phthalocyanine. Relevant elements have been labelled.

| **Paracetamol** | | **Percholorocoronene** | | **Copper(II) Phthalocyanine** | |
|---|---|---|---|---|---|
| Element | Atomic% | Element | Atomic% | Element | Atomic% |
| C | 72.82 | C | 92.07 | C | 76.03 |
| N | 8.17 | N | - | N | 15.79 |
| O | 17.42 | O | 4.02 | O | 2.13 |
| Cl | - | Cl | 1.27 | Cl | - |
| Cu | 1.53 | Cu | 2.11 | Cu | 5.85 |

Table S5: Elemental abundancies from EDS of crystals in this study. Each sample exhibits copper and carbon signals due to the TEM grids used, and oxygen is present due to water ice in cryoTEM.

## 7. Instructions for scripts

### 7.1 *GiveMeED*

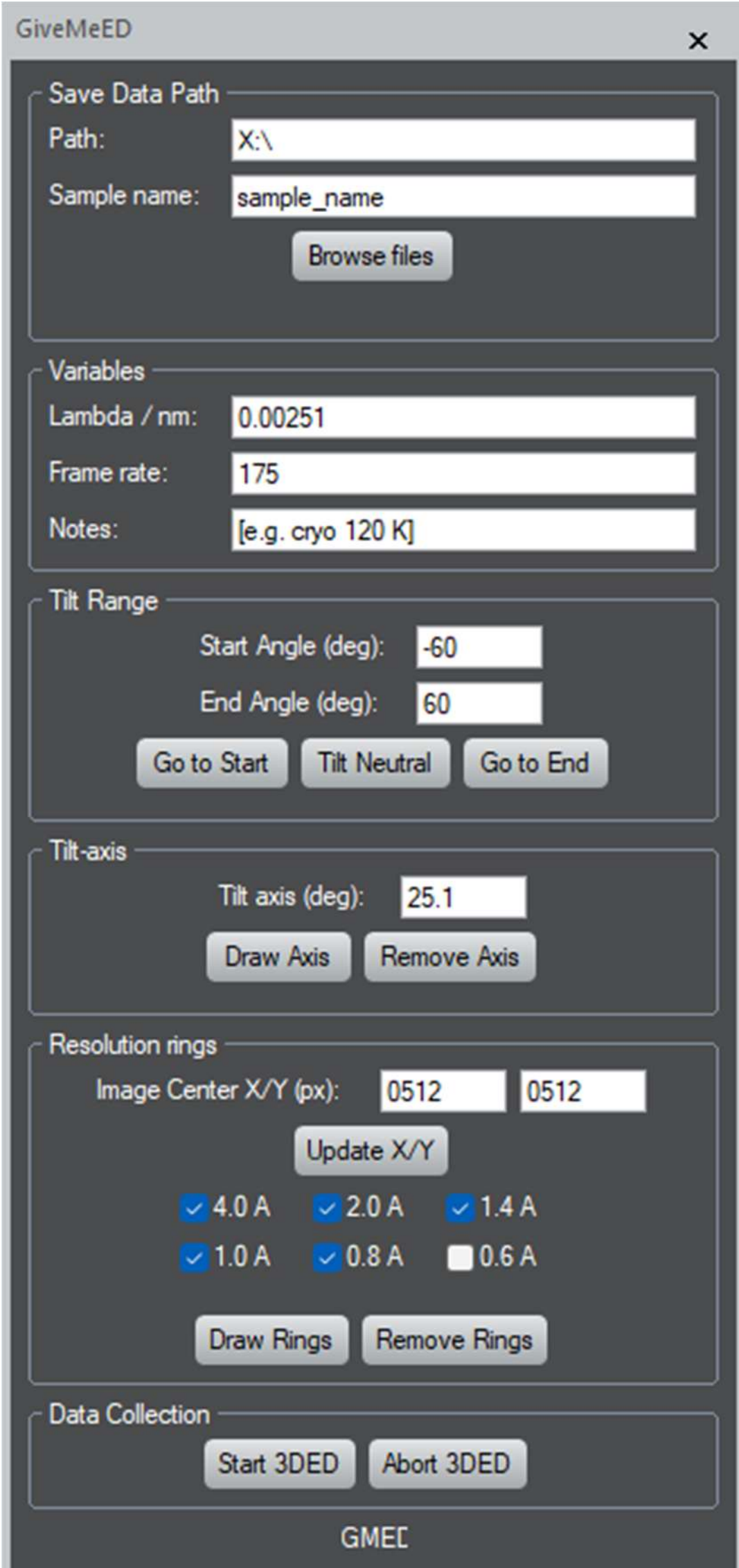

**Figure S4:** *GiveMeED* user interface.

*GiveMeED* has been used successfully with a JEOL 2100Plus TEM with Gatan OneView Camera and JEOL 2100F TEM with Gatan K3 camera. It should work with any TEM using a Gatan camera that has "In-Situ" (video) mode and an installation of *DigitalMicrograph*.

When "Start 3DED" is pressed (Figure S4), *GiveMeED* rotates the TEM stage to the value entered in the "Start Angle" field. The beam blank is turned off and the camera begins recording data while the stage is rotated to the value entered in the "End Angle" field. To prevent the stage colliding with the pole piece, the "End Angle" entered should not exceed the maximum tilt range of the instrument.

Once rotation ends the beam is blanked and all data is saved. Pressing "Abort 3DED" will blank the beam and stop the camera recording. 3DED metadata is saved with the name entered in the "Sample name" field at the path entered in the "Path" field. The values in the "Variables" container are for later reference and do not affect data collection. The recorded metadata is a mixture of information pulled from the microscope during data collection and standard values that can be defined in the script. Modifying the metadata file to include/exclude information can be done by adding/removing lines in the "log_message" string.

The "Abort 3DED" button will blank the beam and stop the camera recording, and the script can also be stopped using the *DigitalMicrograph* kill script shortcut (default "ctrl + numlock"). *GiveMeED* can also be operated without a user interface in which case the variables discussed above are set in the script before executing it. The maximum tilt range of the microscope stage should be checked before attempting 3DED.

Step-by-step instructions for *GiveMeED*:

1. Insert the camera in "In-Situ" mode and start the live view with the frame rate set to a suitable value.
2. In the camera control pane, check the following variables are set correctly: "save data path", and "experiment name". Ensure that the experiment has a unique name. In the *GiveMeED* UI the "Path" field should match "save data path", and the "Sample Name" field should match the "experiment name".
3. Confirm that the specimen is at eucentric height and remains visible over the tilt range entered into the "Tilt Range" container. Pressing the "Go to Start" and "Go to End" buttons will rotate the stage to the value in the corresponding box. "Tilt Neutral" sets the stage angle to 0 degrees.
4. Put the microscope into diffraction mode and press "Start 3DED" to perform data collection. Data and metadata is automatically saved to the file path in Step 2.

The "Tilt-axis" and "Resolution rings" boxes contain the same functionality as *AutoResolutionRings*, described below.

### 7.2 *ExportInSitu*

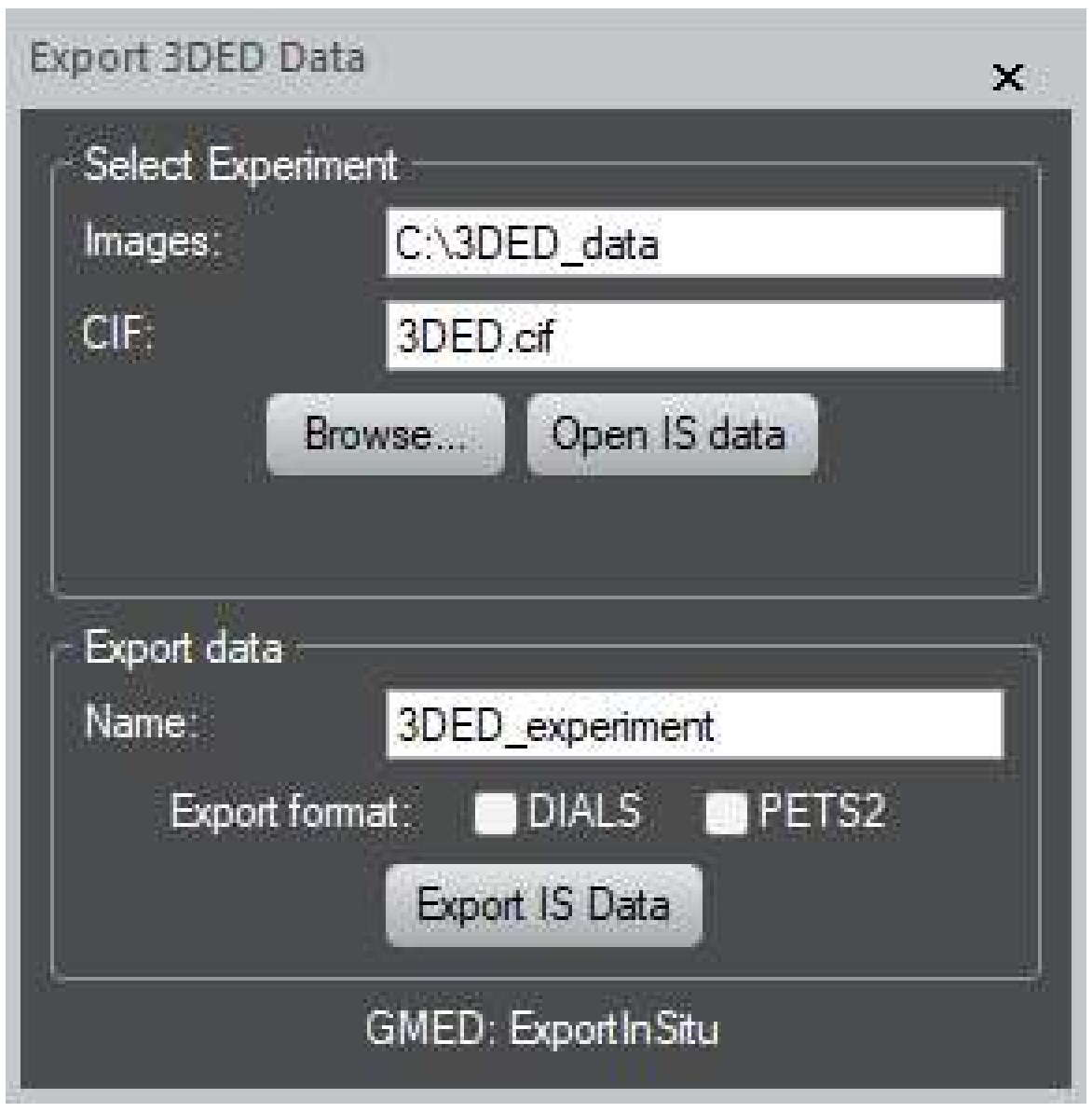

**Figure S5:** *ExportInSitu* user interface.

*ExportInSitu* is used to convert as-recorded 3DED data to a format suitable for data reduction, using the included Python module *export_insitu*. The 3DED dataset is opened as a stack, allowing post-capture processing such as binning, which can then be exported in a format suitable for *DIALS* or *PETS2*. *ExportInSitu* requires the module *export_insitu* to be installed to the *DigitalMicrograph* Python environment.

### 7.2 *Go2Alpha*

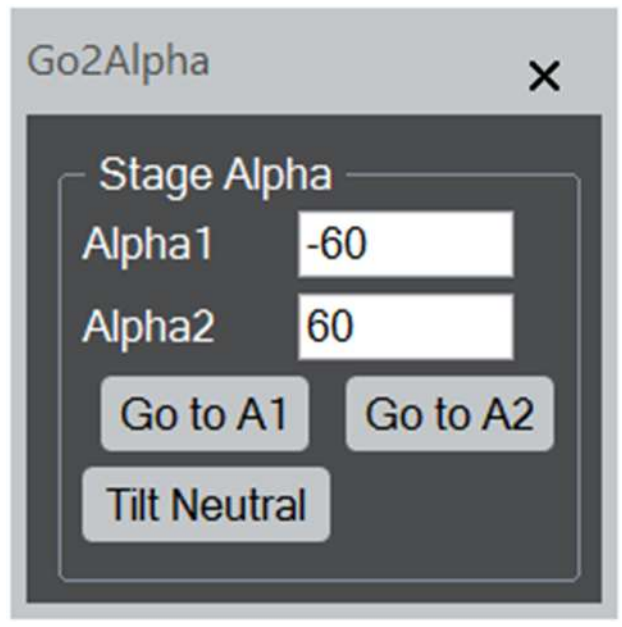

**Figure S6:** *Go2Alpha* user interface.

*Go2Alpha* is used to quickly tilt the stage between two angles (Figure S5). It has the same functionality as the “Tilt Range” container in the *GiveMeED* UI. Two fields can be set: “Alpha1” and “Alpha2” are angles in degrees, where the “Go to A1” button rotates the stage to the angle in the “Alpha1” field (and *vice versa* for “Alpha2” and “Go to A2”). The “Tilt Neutral” button sets the stage angle to 0 degrees.

### 7.3 *AutoResRings*

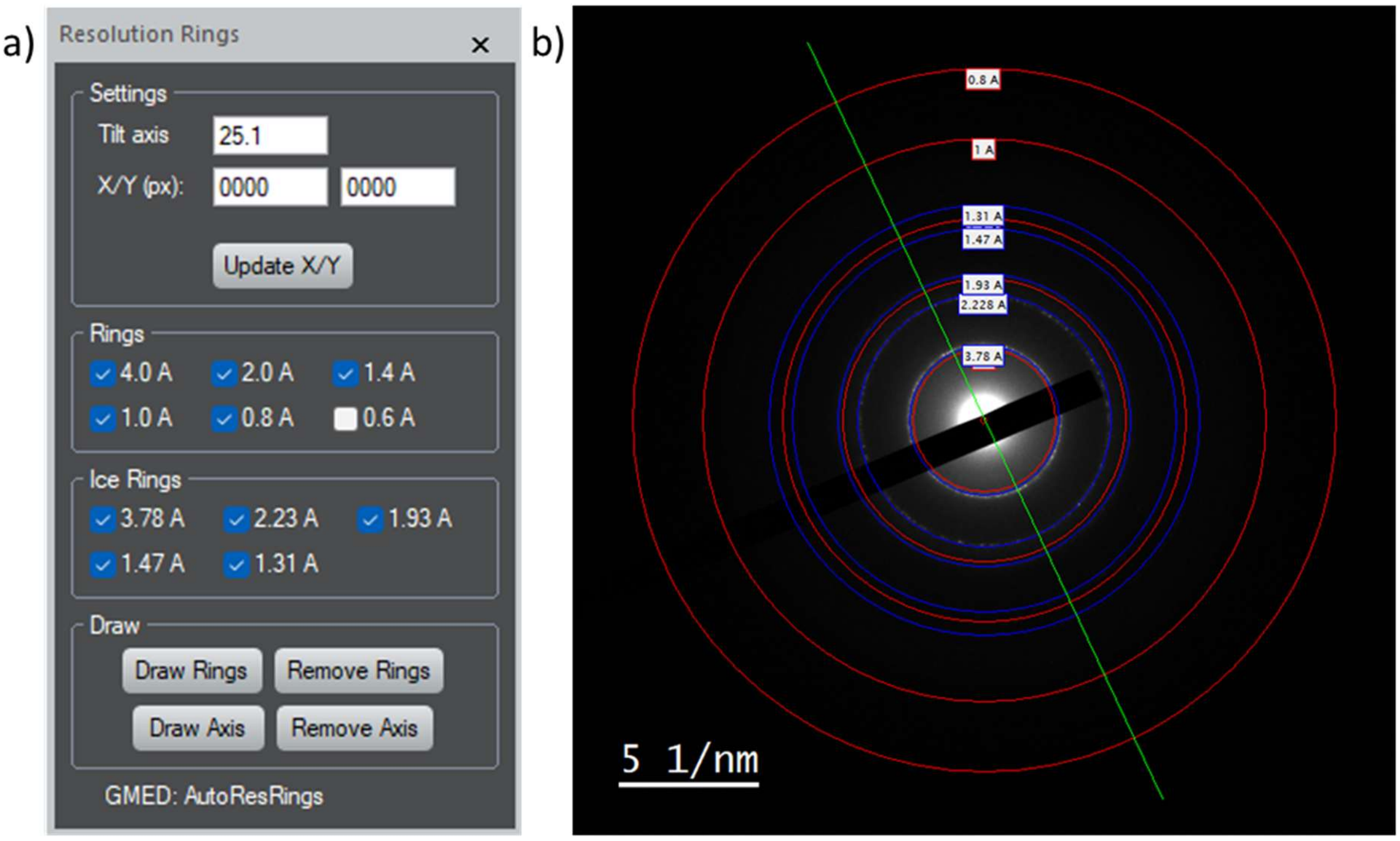

**Figure S7:** (a) *AutoResRings* user interface. (b) Diffraction pattern of ice rings with rings added, using *AutoResRings*, for resolution measurement (red circles), ice rings identification (blue circles), and the tilt axis (green line).

*AutoResRings* is used to draw a set of resolution rings, or the tilt axis, on an image in Digital Micrograph. The tilt axis must be defined in degrees. By default "X/Y" is the geometric centre of the image, but can be adjusted by changing the values in the boxes or pressing the "Update X/Y" button. Rings are drawn on whichever image was last selected, including the camera live view.

By default rings are drawn at 0.8, 1, 1.4, 2, 4 and 100 Å, but different ring radii can be defined in the script. The tick boxes are used to select which rings are drawn, including some common ice ring radii. Scaling the rings is done at creation, so the live view should be in diffraction mode for correct calibration (and the rings will not change size if the camera length is changed). Rings can be deleted from an image by clicking on them and pressing the "delete" key, or using the "Remove Rings" button on the supplied user interface.

Similarly, the tilt axis may be drawn on the image using the "Draw Axis" button and removed with "Remove Axis". The angle is defined in the "Tilt axis" box, in degrees.

## 8. List of Abbreviations

| Abbreviation | Full |
|---|---|
| 3DED | Three-dimensional electron diffraction |
| CCD | Charge-coupled device |
| CIF | Crystallographic information file |
| CMOS | Complementary metal-oxide semiconductor |
| CRED | Continuous rotation electron diffraction |
| CuPC | Copper(II) phthalocyanine |
| ED | Electron diffraction |
| EDS | Energy dispersive X-ray spectroscopy |
| EELS | Electron energy loss spectroscopy |
| PCC | Perchlorocoronene |
| PEDT | Precession-assisted electron diffraction tomography |
| PXRD | Powder X-ray diffraction |
| RED | Rotation electron diffraction |
| SAED | Selected area electron diffraction |
| SCXRD | Single-crystal X-ray diffraction |
| TEM | Transmission electron microscope |
| UI | User interface |

**Table S6:** Abbreviations used in this work.